\documentclass[epj]{svjour}

\usepackage{epsfig}
\usepackage{xspace}
\usepackage{amsmath}
\usepackage{amssymb}

\let\onlinecite\cite

\newcommand{\bk}{\vec{k}}

\newcommand{\sgn}{\mathop{\rm sgn}\nolimits}
\newcommand{\BSCCO}{Bi$_2$\-Sr$_2$\-Ca\-Cu$_2$\-O$_{8+\delta}$\xspace}
\newcommand{\XCOC}{X$_2$\-Cu\-O$_2$\-Cl$_2$\xspace}

\begin{document}

\title{Extended $d_{x^2 - y^2}$-wave superconductivity}
\subtitle{Flat nodes in the gap and the low-temperature asymptotic
   properties of high-$T_c$ superconductors} 
\author{Giuseppe G. N. Angilella\inst{1} \and Asle Sudb\o\inst{2} \and 
   Renato Pucci\inst{1}}
\institute{Dipartimento di Fisica dell'Universit\`a di Catania
   and Istituto Nazionale di Fisica della Materia, U. d. R.
   di Catania,\\ 
   57, Corso Italia, I-95129 Catania, Italy, EU \and
   Institutt for Fysikk, Norges Teknisk-Naturvitenskapelige 
   Universitet NTNU,\\
   Sem S\ae landsvei 9, Gl\o shaugen, N-7491 Trondheim, Norway}
\date{Received: November 19, 1999 / Revised version: December 20, 1999}

\abstract{%
Remarkable anisotropic structures have been recently observed
   in the order parameter $\Delta_\bk$ of the underdoped superconductor
   \BSCCO.
Such findings are strongly suggestive of deviations from a
   simple $d_{x^2 - y^2}$-wave picture of high-$T_c$
   superconductivity, \emph{i.e.} $\Delta_\bk \sim \cos k_x - \cos k_y$.
In particular, flatter nodes in $\Delta_\bk$ are observed along the 
   $k_x = \pm k_y$ directions in $\bk$-space, than within this simple 
   model for a $d$-wave gap.
We argue that nonlinear corrections in the $\bk$-dependence of
   $\Delta_\bk$ near the nodes introduce new energy scales, which
   would lead to deviations in the predicted power-law asymptotic
   behaviour of several measurable quantities, at low or intermediate
   temperatures. 
We evaluate such deviations, either analytically or numerically,
   within the interlayer pair-tunneling model, and within yet another
   phenomenological model for a $d$-wave order parameter.
We find that such deviations are expected to be of different sign in
   the two cases.
Moreover, the doping dependence of the flatness of the gap near the
   nodes is also attributable to Fermi surface effects, in addition to 
   possible screening effects modifying the in-plane pairing kernel,
   as recently proposed.
}

\maketitle

\section{Introduction}

Power laws in the low-temperature asymptotic behaviour of several
   linear response electronic properties provide
   complementary evidence for $d$-wave symmetry of the order parameter 
   (OP) $\Delta_\bk$ of high-$T_c$
   superconductors~\cite{Annett:90,Annett:96} as well as preliminary
   evidence of `exotic' shapes in the OP of heavy fermion
   superconductors, such as UPt$_3$~\cite{Sigrist:91}.
This has to be contrasted to an ``activated'' behaviour $\propto\exp
   (-\beta\Delta_{\rm min} )$, appropriate of $s$-wave superconductors,
   or, in the case of mixed symmetry,
   of superconductors with a non-vanishing $s$-wave contribution to
   their OP, where $\Delta_{\rm min} = \min_\bk |\Delta_{\bk} | > 0$.
In the case of a non-empty nodal manifold for the superconducting
   excitation spectrum $E_\bk$, defined as the locus of points in
   $\bk$-space such that $E_\bk =0$, a large number of quasiparticles
   can be created near such nodes, thus dominating all the low-temperature 
   electronic properties~\cite{Lee:97}.
An exact analysis allows one to relate the exponent of the leading
   power of the low-$T$ expansion of a given linear response function
   to the dimension of the Fermi manifold (defined as the locus of
   states in $\bk$-space with vanishing dispersion relative to the
   Fermi level in the normal state, $\xi_\bk =0$) and the topological
   nature of the nodal manifold, \emph{viz.} a collection of points, of line
   segments, or of surface patches~\cite{Volovik:96,Annett:90}.

On the basis of group theoretical arguments, the simplest choice for a 
   $d$-wave gap function on a square lattice is $\Delta_\bk = \Delta
   g(\bk)$, where $\Delta$ is a $T$-dependent parameter, and
\begin{equation}
g(\bk) = \frac{1}{2} (\cos k_x - \cos k_y )
\label{eq:g}
\end{equation}
is the first basis function associated with the $d$-wave irreducible
   representation of the appropriate crystal point group,
   $C_{4v}$~\cite{Annett:90}. 
We remark that $g(\bk)$ is generated, together with an extended
   $s$-wave term proportional to
\begin{equation}
h(\bk) = \frac{1}{2} (\cos k_x + \cos k_y ), 
\label{eq:h}
\end{equation}
by a nearest-neighbour interaction term in real space.
Here and in the following we shall measure the wavevectors in units of 
   the appropriate inverse lattice spacings.
Proportionality to Eq.~(\ref{eq:g}) allows $\Delta_\bk$ to vanish
   linearly at a given point along the Fermi line, which for most cuprate
   superconductors can be modelled by the tight-binding expansion:
\begin{equation}
\xi_\bk = -2t(\cos k_x + \cos k_y ) + 4t^\prime \cos k_x \cos k_y -
   \mu = 0,
\label{eq:disp}
\end{equation}
where $t=0.25$~eV, $t^\prime = 0.45 t$ measure nearest and
   next-nearest neighbour hopping, respectively, and $\mu$ is the
   chemical potential.

On the other hand, increasing experimental evidence above all from
   angle-resolved photoemission spectroscopy (ARPES) suggests a richer 
   structure in $\bk$-space for the OP of the underdoped high-$T_c$
   superconductor \BSCCO~\cite{Mesot:99}.
In particular, the superconducting gap near the nodal points turns
   out to be flatter than predicted by the simple assumption
   $\Delta_\bk \propto g(\bk)$~\cite{Mesot:99}.
Such a feature is consistent with the observation of whole ungapped
   segments of the Fermi line above $T_c$ in the pseudogap regime of
   underdoped \BSCCO~\cite{Norman:98}, and will of course serve as a
   constraint for a microscopic understanding of the pairing
   mechanism.

Quite remarkably, qualitatively similar deviations from a
   $g(\bk)$-like dispersion have been evidenced in the
   $\bk$-depend\-ence of the antiferromagnetic gap in the related
   insulating compounds \XCOC (X = Ca, Sr) \cite{Ronning:98}.
Such a finding has been interpreted in terms of an interrelation
   between the antiferromagnetic phase of the parent insulator and the
   underdoped regime of the intervening superconductor \cite{Zacher:99}.

In this paper, we argue that such extended structures in the
   superconducting OP, interpolating between point and line nodes, can
   be included in the definition of $\Delta_\bk$ as higher order terms
   in $g(\bk)$.
We shall then look for their signatures in the low-temperature
   asymptotic electronic properties of the superconducting cuprates,
   as corrections to the predicted power-law behaviour.
In deriving our results analytically, we will specifically consider
   the interlayer pair-tunneling (ILT) mechanism of high-$T_c$
   superconductivity~\cite{Chakravarty:93}, which has been shown to
   accurately reproduce most of the observed gap
   features~\cite{Angilella:99}. 

\section{Extended $d$-wave gap within the ILT model}

A distinguishing feature of the ILT mechanism, compared to other
   proposed models of HTSC, is that superconductivity is driven by a
   gain in kinetic, rather than potential, energy as temperature is
   lowered below the critical temperature $T_c$.
It is assumed that coherent single particle hopping between adjacent
   CuO$_2$ layers in the cuprates is suppressed by the non-Fermi
   liquid character of the normal state (\emph{e.g.} due to spin-charge
   separation), while interlayer coherent tunneling of Cooper pairs is
   allowed as soon as a superconducting condensate is established.
Confined coherence~\cite{Clarke:97} within
   CuO$_2$ layers in the 
   normal state is indeed largely motivated by the absence of coherent 
   transport along the $c$-axis, whereas a comprehensive theoretical
   understanding of it is still lacking.
However, there is now abundant \emph{experimental} evidence that
   $c$-axis transport in the normal state indeed is incoherent, while
   that in the superconducting state may not be~\cite{vanderMarel}.
This seems to warrant attention being paid to unconventional models of 
   high-$T_c$ superconductivity based on relieving $c$-axis frustrated 
   kinetic energy.
Recent findings~\cite{Moler:98,Tsvetkov:98} suggest,
   however, that the ILT mechanism alone is not sufficient to account
   for the large condensation energy $E_c$, as extracted
   experimentally from measurements of the penetration length $\lambda_c$
   of several single layered compounds, such as
   Tl$_2$Ba$_2$CuO$_{6+\delta}$~\cite{Moler:98,Tsvetkov:98},
   whereas the predictions of the ILT model agrees with the measured
   value of $E_c$ for
   La$_{2-x}$Sr$_x$CuO$_4$~\cite{Panagopoulos:97,Panagopoulos:99,Leggett:98,Anderson:98}.
It has been pointed out, however, that while considerable experimental 
   effort has
   been devoted to the determination of $\lambda_c$, extracting $E_c$
   from existing data on electronic specific heat is by no means
   straightforward~\cite{Chakravarty:99}.
A direct evaluation of $E_c$ from its mean-field expression at $T=0$
   \cite{Schrieffer:64} would relieve the complications arising from
   thermal fluctuation effects, inherent in the method of integrating
   specific heat data, from $T=0$ through $T_c$, recently pointed out
   in Ref.~\onlinecite{Chakravarty:99}.
By utilizing the gap equation, Eq.~(\ref{eq:delta}), within the ILT
   model, and of the expression relating $\lambda_c$ at $T=0$ to $E_c$ 
   \cite{Chakravarty:98}, we find results for $\lambda_c (T=0)$ in
   Bi2212 which are within factors of order unity from the
   experimental values, rather than factors of 10 to 20
   \cite{Angilella:00}.
The observed doping dependence of $\lambda_c$ \cite{Panagopoulos:99}
   is also qualitatively reproduced \cite{Angilella:00}.

The emerging scenario suggests therefore that some in-plane effective
   interaction might co-operate with the ILT mechanism in establishing 
   the superconducting state~\cite{note:int-kin}.
One could think of such a mechanism as a seed for the Cooper
   instability, and the origin of the gap's dominant $d$-wave symmetry.
Once Cooper pairs are formed in the appropriate symmetry channel(s)
   via such in-plane effective interaction, the ILT mechanism would
   allow the condensate for an additional energy gain, by releasing
   the constraint of in-plane segregation.

Without explicitly specifying the microscopic origin of the in-plane
   mechanism, we therefore assume the in-plane pairing potential to
   be given by $V_{\bk\bk^\prime} = V g(\bk) g(\bk^\prime )$ ($V <0$),
   thus allowing for $d$-wave symmetry of the order parameter. 
The issue of the competition with other subdominant ($s$-wave)
   symmetry channels in the presence of ILT has been addressed in
   Ref.~\onlinecite{Angilella:99}, showing that the $d$-wave
   contribution wins out at optimal doping and in the underdoped
   regime. 
Despite its kinetic nature, ILT can be absorbed in the interacting
   part of the Hamiltonian as an effective term 
   $T_J (\bk)\delta_{\bk\bk^\prime}$, whose $\bk$-space locality
   enforces in-plane momentum conservation during a tunneling
   process~\cite{Chakravarty:93}.
Following Ref.~\onlinecite{Chakravarty:93}, we assume $T_J (\bk) =
   t_\perp^2 (\bk)/t \equiv T_J g^4 (\bk)$, being $t_\perp (\bk)$ the
   single-particle interlayer hopping amplitude, with $t_\perp (\bk)
   \propto g^2 (\bk)$, as suggested by ARPES as well as by band
   structure calculations \cite{Chakravarty:93,Andersen:96}.
A standard mean-field diagonalization technique then yields the
   following expression for the energy
   gap~\cite{Sudbo:95,Angilella:99}:
\begin{equation}
\Delta_\bk = \frac{\Delta g(\bk)}{1-T_J (\bk)\chi_\bk} ,
\label{eq:delta}
\end{equation}
where $\chi_\bk = (2E_\bk )^{-1} \tanh (\beta E_\bk /2)$ is the
   superconducting pair susceptibility, and $E_\bk = (\xi_\bk^2 +
   |\Delta_\bk |^2 )^{1/2}$ is the upper branch of the superconducting 
   elementary excitation spectrum.

Along the Fermi line ($\xi_\bk =0$) at $T=0$, one immediately sees
   that:
\begin{equation}
\Delta_\bk = \Delta g(\bk) + \frac{1}{2} T_J g^4 (\bk) \sgn
   [g(\bk)].
\label{eq:deltaFL}
\end{equation}
Such an expression, together with manifestly fulfilling the
   requirement of $d$-wave symmetry, also endows the superconducting
   gap with a richer structure near the nodal points along the $k_x =
   \pm k_y$ directions.
This is probably best seen by considering the Fourier expansions:
\begin{subequations}
\label{eqs:expansion}
\begin{eqnarray}
\label{eq:expansion-g}
g(\bk) &&= -2\sum_{m=1}^\infty J_{4m-2} (k) \cos[(4m-2)\phi],\\
\label{eq:expansion-h}
h(\bk) &&= J_0 (k) + 2 \sum_{m=1}^\infty J_{4m} (k) \cos(4m\phi),\\
\label{eq:expansion-TJ}
T_J (\bk)/T_J &&= \frac{9}{64} + \frac{a_0}{2} +
   \sum_{m=1}^\infty a_{4m} \cos (4m\phi),
\end{eqnarray}
\end{subequations}
with
\begin{eqnarray}
a_{4m} &&= \frac{1}{32} J_{4m} (4k) + \frac{1}{2} J_{4m} (2k) +
   \frac{3}{16} (-1)^m J_{4m} (2k\sqrt{2}) \nonumber \\
&& - \frac{3}{4} (-1)^m J_{4m} (k\sqrt{2}) -\frac{1}{4} J_{4m}
   \left( \frac{k}{\sin\phi_0} \right) \cos(4m\phi_0 ). \nonumber \\
\end{eqnarray}
Here, the generic wavevector $\bk$ is expressed in terms of its
   modulus $k$ and of the angle $\phi$ formed with the $\Gamma X$
   direction in the first Brillouin zone (1BZ), $\bk =
   (k\cos\phi,k\sin\phi)$, $J_\alpha (x)$ are Bessel functions of the
   first kind and order $\alpha$, and $\tan\phi_0 = \frac{1}{3}$.

\begin{figure}
\centering
\epsfig{width=0.95\columnwidth,angle=-90,file=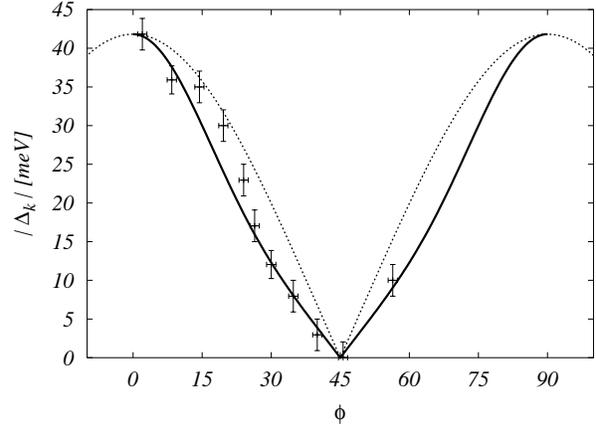}
\caption{%
Fit for $|\Delta_{\bf k} |$ within the ILT model,
   Eq.~(\protect\ref{eq:deltaFL}) (solid line), and in the case of a
   simple $d$-wave gap $|\Delta_{\bf k}| = \Delta g({\bf k})$
   (dashed line), against Mesot {\em et al.}'s ARPES data for
   underdoped Bi2212 ($T_c = 75$~K, Ref.~\protect\cite{Mesot:99}).}
\label{fig:fit}
\end{figure}

Eq.~(\ref{eq:deltaFL}) is to be contrasted to the phenomenological fit
\begin{equation}
\Delta_\bk = \Delta [B\cos(2\phi) +(1-B)\cos(6\phi)]
\label{eq:Mesot}
\end{equation}
proposed in Ref.~\onlinecite{Mesot:99} for $\Delta_\bk$ along the
   Fermi line: 
Instead of requiring an in-plane interaction extended to further
   neighbours, Eq.~(\ref{eq:delta}) endows the superconducting gap with 
   the observed flat structure around the nodes, through the ILT term
   $T_J (\bk)$.
In Fig.~\ref{fig:fit}, we fit Eq.~(\ref{eq:deltaFL}) against Mesot {\em
   et al.}'s experimental data for one of the underdoped Bi2212
   samples in Ref.~\cite{Mesot:99}, having $T_c = 75$~K. 
A remarkable agreement follows already by fixing $\Delta$ so that
   $|\Delta_{\bf k} |$ reproduces the maximum datum at $\bk = (\pi,\pi)$,
   whereas $T_J$ is taken to be 0.04~eV \cite{Chakravarty:93}.
In particular, besides obtaining an enhanced maximum value of $|\Delta_{\bf k}
   |$ at $\bk=(\pi,\pi)$, we are thus able to recover the anomalously
   flat region around the node at $\phi=45^\circ$ in a rather natural way.
We note, however, that our fit requires $\Delta\approx T_J /2$ around
   optimal doping, which will not be without consequences in
   evaluating other fundamental quantities \cite{Angilella:00}. 

Eq.~(\ref{eq:deltaFL}) already contains the doping dependence of the
   observed gap anisotropy, although in a hidden way.
As pointed out in Ref.~\onlinecite{Angilella:99}, the auxiliary parameter
   $\Delta$ is to be self-consistently determined by solving the
   appropriate gap equation.
Besides being intrinsically doping dependent, this equation is
   unconventionally modified by the presence of a $\bk$-local
   effective interaction, as induced by the ILT mechanism.
Moreover, the role of the contribution $\propto T_J g^4 (\bk)$ in
   Eq.~(\ref{eq:deltaFL}) is strongly influenced by the actual
   location of the Fermi line, as $g^4 (\bk)$ is sharply peaked at
   $\bk = (0,\pi)$ (and symmetry related points).

Eq.~(\ref{eq:deltaFL}) also facilitates the evaluation of the slope of the
   superconducting gap $v_\Delta = (1/2) d|\Delta_\bk |/d\phi$ at the
   nodal point along the Fermi line.
Such a quantity is related to the temperature derivative of the
   superfluid stiffness at $T=0$.
In particular, it is seen that the ratio $v_\Delta /\Delta_{\rm max}$ 
   decreases with underdoping~\cite{Mesot:99}.
From Eq.~(\ref{eq:deltaFL}), one derives that
   $v_\Delta$ is independent of $T_J$, and that therefore a doping induced
   change of $v_\Delta$ through $\Delta$ essentially can be traced
   back to the actual position of the Fermi line, as discussed above,
   within the ILT model.
The ratio $v_\Delta / \Delta_{\rm max}$ will anyway deviate from its
   value within simple $d$-wave (BCS-like) models, as a function of doping,
   due to the enhancement of $\Delta_{\rm max}$ induced by ILT.

\section{Low-temperature asymptotic behaviour of electronic properties}

We now address the issue, whether such extended features of the OP near 
   the nodes, as those described in the previous section, induce 
   deviations in the low or intermediate temperature asymptotic
   behaviour of linear response electronic properties in the
   superconducting state.
In what follows, we shall limit our discussion to clean
   superconductors, and neglect impurity effects altogether.
Mean-field (BCS or BCS-like) expressions for most linear response
   electronic properties are available also in the case of
   anisotropic, \emph{i.e.} non $s$-wave, superconductors.
In particular, we have in mind observable quantities such as the
   superconducting density~\cite{Leggett:75}, the electronic
   specific heat~\cite{Leggett:75}, the spin
   susceptibility~\cite{Leggett:75}, the penetration
   depth~\cite{Scalapino:92}, the thermal
   conductivity~\cite{Bardeen:59}, and so on.
Their expressions basically involve the evaluation of some integral of
   the kind:
\begin{equation}
{\cal F}[\beta; \varphi_\bk (\beta)] = \frac{1}{(2\pi)^2} \int d^2 \bk 
   \varphi_\bk (\beta) e^{-\beta E_\bk} ,
\label{eq:integral}
\end{equation}
where $\beta=(k_{\rm B} T)^{-1}$, $\varphi_\bk (\beta)$ is a
   (dimensional) function of wavevector $\bk$ and temperature, related 
   to the electronic quantity of interest, and the integration is
   extended to the 1BZ, $\bk\in[-\pi,\pi]\times[-\pi,\pi]$ (see
   App.~\ref{app:electronic}). 
In the case of $d$-wave superconductors, $E_\bk$ is allowed to vanish
   at the intersection between the Fermi line and the nodal lines of
   the gap function.
Around such points, quasiparticles can be created in large numbers.
In the limit of low temperatures ($\beta\to\infty$), therefore, the
   value of the integral in Eq.~(\ref{eq:integral}) is dominated by
   the contributions from wavevectors $\bk$ close to such point nodes.
Around such nodes, it is useful to introduce the new sets of
   coordinates $(k_1 ,k_2 )$ or $(\epsilon,\theta)$, defined
   as~\cite{Lee:97}:
\begin{subequations}
\label{eq:conecoords}
\begin{eqnarray}
\xi_\bk &&\simeq {\bf v}_{\rm F} \cdot \bk \equiv v_{\rm F} k_1 =
   \epsilon\cos\theta , \\
\Delta g(\bk) &&\simeq {\bf v}_2 \cdot \bk \equiv v_2 k_2 =
   \epsilon\sin\theta,
\end{eqnarray}
\end{subequations}
in units where $\hbar=1$.
Here, $v_{\rm F}$ and $v_2$ are the Fermi velocity and a suitable
   `gap' velocity, respectively, evaluated at $E_\bk =0$, and
   $\epsilon$ measures the distance in energy from a given dispersionless
   point implicitly defined by $E_\bk =0$. 
In terms of the new coordinates, the superconducting spectrum for a
   simple $d$-wave superconductor near a node therefore looks
   like an anisotropic Dirac cone~\cite{Lee:97},
\begin{equation}
E_\bk \sim \left(v_{\rm F}^2 k_1^2 + v_2^2 k_2^2 \right)^{1/2} =\epsilon.
\label{eq:Dirac}
\end{equation}

\begin{figure}
\centering
   \epsfig{file=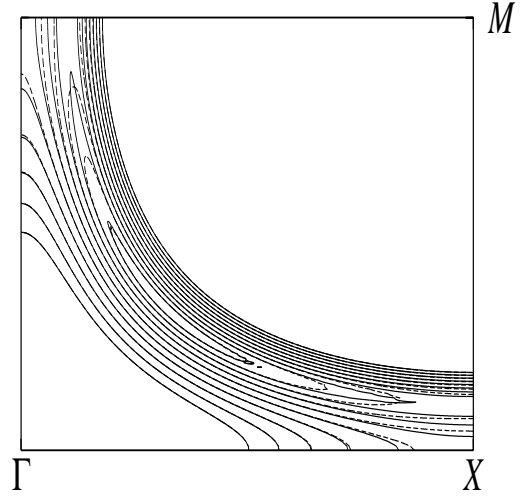,height=0.8\columnwidth,width=0.8\columnwidth,angle=-90}
\caption{Typical contour lines of the superconducting spectrum $E_\bk$ in the
   simple $d$-wave case (dashed lines) and in presence of ILT
   (continuous line).}
\label{fig:dispersion}
\end{figure}

\begin{figure}
\centering 
   \epsfig{file=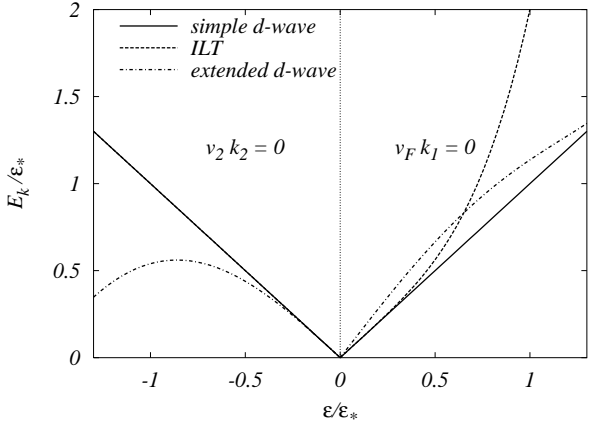,width=0.95\columnwidth,angle=-90}
\caption{Deviations from the simple $d$-wave case,
   Eq.~(\protect\ref{eq:Dirac}) (continuous line), of the
   superconducting spectrum $E_\bk$ around a node, within the ILT
   model, Eq.~(\protect\ref{eq:EkILT}) (dashed line), and in the case
   of an extended $d$-wave gap, Eq.~(\protect\ref{eq:Ekext})
   (dashed-dotted line), as a function of the reduced coordinates
   $\epsilon/\epsilon_\star$ ($\epsilon/\tilde\epsilon_\star$,
   respectively). 
Such deviations are most easily seen along the direction of the nodal
   line (left panel, $v_2 k_2 =0$ or $\theta=\pi$), and along the Fermi
   line (right panel, $v_{\rm F} k_1 = 0$ or $\theta=\pi/2$).
Note that along $v_2 k_2 =0$, one has $E_\bk =\epsilon$ also within
   the ILT model.
}
\label{fig:cones}
\end{figure}

The observation of flatter structures near the nodes~\cite{Mesot:99}
   not only implies a more significant anisotropy ratio $v_{\rm F} /
   v_2$, but also the possibility that higher order terms in
   $\epsilon$ may contribute to $E_\bk$, Eq.~(\ref{eq:Dirac}).
Indeed, within the ILT model, from Eq.~(\ref{eq:delta}) at $T=0$ one obtains
\begin{equation}
E_\bk \sim \epsilon \left[1 + \left(\frac{\epsilon}{\epsilon_\star}
   \right)^3 \sin^6 \theta \right], 
\label{eq:EkILT}
\end{equation}
to lowest order in $\epsilon/\epsilon_\star$, with $1/\epsilon_\star^3
   = (1/2)(T_J /\Delta^4 )$ related to the pair-tunneling amplitude
   $T_J$ and to the auxiliary gap parameter $\Delta$
   (Figs.~\ref{fig:dispersion} and \ref{fig:cones}).

Other models, based on extended in-plane
   pairing mechanisms, would in general yield different polynomial
   corrections in $\epsilon$ to $E_\bk$.
For instance, within the spin fluctuation theory~\cite{Monthoux:91},
   the following phenomenological expansion holds for the momentum
   distribution of the superconducting energy gap~\cite{Ghosh:99}
\begin{equation}
\Delta_\bk = \Delta g(\bk) \sum_{n=0}^N d_n h^n (\bk),
\label{eq:SF}
\end{equation}
with all coefficients $d_n =1$.
We explicitly observe that for $N=0$, $d_0 =1$, one recovers the
   simple $d$-wave gap $\Delta_\bk \propto g(\bk)$, while the
   case $N=1$, with the identifications $\Delta \mapsto B\Delta$,
   $d_0 = 1$, $d_1 = 4(1-B)/B$, maps to~\cite{Mesot:99,Ghosh:99}:
\begin{equation}
\Delta_\bk = \Delta [ Bg(\bk) + (1-B)g(2\bk)],
\label{eq:Mesot2}
\end{equation}
which is compatible with the phenomenological fit Eq.~(\ref{eq:Mesot}) 
   proposed by Mesot \emph{et al.} in Ref.~\onlinecite{Mesot:99} for
   their experimental data of $|\Delta_\bk |$ along the Fermi
   line~\cite{Mesot:99,Ghosh:99}. 
In particular, Eq.~(\ref{eq:Mesot2}) would follow from a correction
   $\delta V_{\bk\bk^\prime} \propto g(2\bk)g(2\bk^\prime )$ to the
   in-plane coupling, corresponding to next-nearest neighbours
   interaction.

In such a particular case, and
   assuming for simplicity $t^\prime = \mu =0$ in Eq.~(\ref{eq:disp}),
   one straightforwadly obtains 
\begin{equation}
E_\bk \sim \epsilon \left[ \cos^2 \theta + \sin^2 \theta \left( 1 -
   \frac{\epsilon}{\tilde\epsilon_\star} \cos\theta \right)^2
   \right]^{1/2} ,
\label{eq:Ekext}
\end{equation}
where $\tilde\epsilon_\star = tB/(1-B)$ is now related to the ratio of 
   nearest \emph{vs} next-nearest neighbours coupling.
Therefore, both within the ILT model and within other models, based on 
   extended in-plane pairing, the additional mechanism responsible for 
   the nonlinear correction to $E_\bk$ away from its nodes introduces
   new energy scales (here, $\epsilon_\star$ or
   $\tilde\epsilon_\star$, respectively). 
Fig.~\ref{fig:cones} depicts the two different ways in which $E_\bk$
   deviates from the cone-like shape, Eq.~(\ref{eq:Dirac}), near a
   node, in the two cases given by Eqs.~(\ref{eq:EkILT}) and
   (\ref{eq:Ekext}).

In the absence of any such additional mechanism ($\epsilon_\star,
   \tilde\epsilon_\star = 0$), the leading contribution to
   Eq.~(\ref{eq:integral}) for the simplest, reference case $\varphi_\bk
   (\beta)\equiv 1$ is:
\begin{equation}
{\cal F}_1 (\beta) \equiv {\cal F}[\beta;1] \doteq \frac{A}{\beta^2} ,
\label{eq:conventional}
\end{equation}
where $A = (2\pi v_{\rm F} v_2 \Delta )^{-1}$ is a doping-dependent
   factor, and $\doteq$ denotes equality up to terms vanishing
   exponentially with $\beta$ at \emph{all} energy scales, as
   $\beta\to\infty$ ($T\to0$).
Eq.~(\ref{eq:conventional}) should be regarded as typical of the
   power-law asymptotic low-temperature behaviour of the superconducting
   electronic properties within a simple $d$-wave BCS-like model. 

\begin{figure}
\centering 
   \epsfig{file=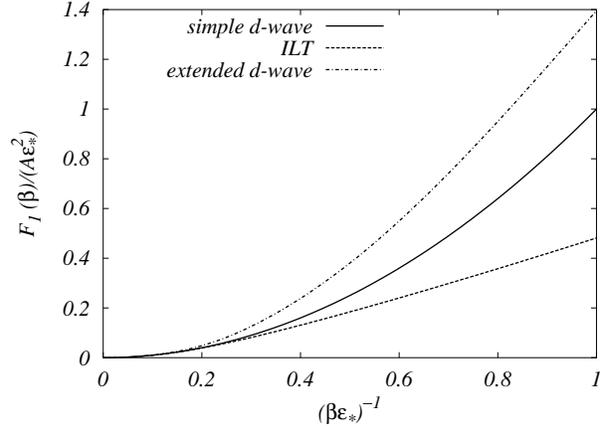,width=0.95\columnwidth,angle=-90}
\caption{Asymptotic power-law ($\propto T^2$, solid line) and modified
   power-law behaviours of ${\cal F}_1 (\beta)$, in the presence of ILT
   (dashed line), and in the extended $d$-wave case (dashed-dotted
   line), as $(\beta\epsilon_\star )^{-1} \to0$.
Given the values used for the fits of $|\Delta_\bk |$ along the Fermi
   line in Refs.~\protect\onlinecite{Mesot:99} and
   \protect\onlinecite{Angilella:99}, it turns out that
   $\epsilon_\star \sim 250$~K.}
\label{fig:comparison}
\end{figure}

In order to obtain an asymptotic expansion for ${\cal F}_1 (\beta)$ as
   $\beta\to\infty$ ($T\to0$), including the corrections due to ILT,
   Eq.~(\ref{eq:EkILT}), we observe that the integration over
   $\epsilon$ in Eq.~(\ref{eq:integral}) is actually made of two
   contributions:
\begin{equation}
\int_0^\infty d\epsilon = \int_0^{\epsilon_\star} d\epsilon +
   \int_{\epsilon_\star}^\infty d\epsilon.
\end{equation}
In the first integral, we may safely retain only the linear term $E_\bk
   \sim \epsilon$ in the exponent, since $\epsilon\leq\epsilon_\star$.
In the second contribution, this is no longer possible, and
   Eq.~(\ref{eq:EkILT}) has to be retained in full.
However, since $\epsilon\geq\epsilon_\star > 0$, one can make use of
   Laplace's (saddle point) method for the integral over angles around 
   $\theta=0$.
The final result is:
\begin{eqnarray}
{\cal F}_1 (\beta) &&\sim \frac{A}{\beta^2} \left[ 1 -
   (1+\beta\epsilon_\star ) e^{-\beta\epsilon_\star} + \frac{1}{3\pi}
   \Gamma \left( \frac{1}{6} \right) \beta\epsilon_\star \Gamma \left( 
   \frac{4}{3} , \beta\epsilon_\star \right) \right] \nonumber \\
&&\sim \frac{A}{\beta^2} \left[ 1 - \left( 1 + \beta\epsilon_\star -
   \frac{1}{3\pi} \Gamma \left( \frac{1}{6} \right)
   (\beta\epsilon_\star )^{5/6} \right) e^{-\beta\epsilon_\star}
   \right],
\label{eq:asymptotic}
\end{eqnarray}
where $\Gamma(x)$, $\Gamma(\alpha,x)$ are Euler gamma and incomplete
   gamma functions, respectively~\cite{Gradshteyn:94}.
A comparison of Eq.~(\ref{eq:conventional}) and
   Eq.~(\ref{eq:asymptotic}) is provided by Fig.~\ref{fig:comparison}, 
   and shows that ${\cal F}_1 (\beta)$ gets effectively \emph{suppressed} 
   in the presence of flat nodes in the order parameter, as provided
   by the ILT mechanism, with respect
   to the simple $d$-wave case, at an energy scale
   $\sim\epsilon_\star$. 

No such simple asymptotic expansion for ${\cal F}_1 (\beta)$ is available 
   in the extended $d$-wave case described by Eq.~(\ref{eq:Ekext}),
   and the integrations have to be performed numerically.
Fig.~\ref{fig:comparison} shows the result, with the identifications
   $A\mapsto \tilde A = (2\pi v_{\rm F} v_2 B\Delta )^{-1}$ and
   $\epsilon_\star \mapsto \tilde\epsilon_\star$.
In this case, Eq.~(\ref{eq:Ekext}) provides $E_\bk$ with a different
   kind of anisotropy with respect to the simple case,
   Eq.~(\ref{eq:Dirac}), than Eq.~(\ref{eq:EkILT}) does.
While in the latter case one always has $E_\bk \geq \epsilon$, here one 
   has $E_\bk \gtreqless\epsilon$, depending on the angle $\theta$
   (cfr. Fig.~\ref{fig:cones}). 
As a consequence, ${\cal F}_1 (\beta)$ is \emph{enhanced}
   with respect to the simple $d$-wave case, at an energy scale
   $\sim\tilde\epsilon_\star$. 

\section{Conclusions}

Motivated by recent experimental findings of extended flat structures
   in the order parameters of the underdoped $d$-wave superconductor
   \BSCCO~\cite{Mesot:99}, we have addressed the issue of whether
   nonlinear, high-energy corrections to the superconducting energy
   spectrum $E_\bk$ around the gap nodes induce deviations in
   the predicted power-law behaviour of several electronic properties
   at low or intermediate temperatures.
We have shown that nonlinear corrections to $E_\bk$
   in general introduce additional energy scales in the problem.
Deviations from the usual power-law behaviour of the
   superconducting electronic properties are indeed to be expected at such
   energy scales, but the actual value and \emph{sign} of such
   deviations are specific to the model under consideration.
In particular, within the ILT model, we have explicitly derived the
   expected corrections to a typical power-law asymptotic behaviour as
   $T\to0$, showing these to be \emph{negative,} whereas within a
   phenomenological model of extended $d$-wave
   superconductivity~\cite{Mesot:99} such corrections are predicted to 
   be \emph{positive.}
Whether such deviations will actually be observable in real
   measurements of superconducting electronic properties, will of
   course depend on the effective values of the additional energy
   scales $\epsilon_\star$ or $\tilde\epsilon_\star$ in real compounds.

\begin{acknowledgement}
G. G. N. A. thanks P. Falsaperla, J. O. Fj\ae restad, and Ch.~W\"alti for 
   valuable discussions, and acknowledges the NTNU (Trondheim, Norway)
   for warm hospitality and financial support during the period in
   which the present work was brought to completion.
A. S. acknowledges support from Norges Forskningsr{\aa}d through
   Grants No.~110566/410 and No.~110569/410.
\end{acknowledgement}

\appendix

\section{Low-temperature superconducting electronic properties}
\label{app:electronic}

We now give a sketch of how the low-temperature asymptotic behaviour of 
   several electronic properties of interest can be reduced to that of 
   ${\cal F}_1 (\beta)$ or its derivatives.
Most electronic quantities in the superconducting state are in fact
   given by Eq.~(\ref{eq:integral}), with $\varphi_\bk (\beta)$
   actually depending on $\bk$ only through $E_\bk$.
In what follows, $f(\epsilon) = [1+\exp(\beta\epsilon)]^{-1}$ denotes
   the Fermi function.

Within BCS theory, the electronic specific heat is given
   by~\cite{Leggett:75}:
\begin{eqnarray}
C_V &&= \sum_\bk 2 k_{\rm B} \beta E_\bk \left[ E_\bk + \frac{\partial 
   E_\bk}{\partial\beta} \right] \left( -\frac{\partial f}{\partial
   E_\bk} \right) \nonumber \\
&&\sim \sum_\bk 2 k_{\rm B} \beta E_\bk^2 \left( -\frac{\partial f}{\partial
   E_\bk} \right) \nonumber\\
&&\sim 2k_{\rm B} \beta^2 {\cal F}_1^{\prime\prime} (\beta),
\end{eqnarray}
where apices denote derivatives with respect to $\beta$.
Here, $\varphi_\bk (\beta) = 2k_{\rm B} \beta^2 E_\bk^2$, and we have
   made use of the fact that $(-\partial f/\partial E_\bk ) \doteq
   \beta\exp(\beta E_\bk)$.

Analogously, the unrenormalized, static, isotropic spin susceptibility 
   $\chi_0 = \chi_0 ({\bf q}\to 0,\omega\to 0)$, which is directly
   related to the Knight shift, is simply given
   by~\cite{Schrieffer:64,Sudbo:94}:
\begin{equation}
\chi_0 = \sum_\bk \left( -\frac{\partial f}{\partial E_\bk} \right)
   \doteq \beta {\cal F}_1 (\beta).
\end{equation}

The expression of the electronic thermal conductivity for an
   anisotropic $d$-wave  
   superconductor also involves an average of $(-\partial f/\partial
   E_\bk )$ over the 1BZ~\cite{Bardeen:59,Krishana:97}:
\begin{equation}
\kappa_e = \frac{1}{T} \sum_\bk \left( -\frac{\partial f}{\partial
   E_\bk} \right) E_\bk^2 \left( \frac{\partial E_\bk}{\partial k_x}
   \right)^2 \tau (\bk),
\end{equation}
where $\tau(\bk)$ is the superconducting quasiparticles lifetime.
Due to the presence of the $x$ component of the group velocity
   $\nabla_\bk E_\bk$, however, its expression in our notation reduces 
   to:
\begin{equation}
\kappa_e = \frac{1}{8\pi^2} \frac{\ell_0}{v_2} \frac{1}{T}
   \int_0^\infty d\epsilon\,\epsilon^3 \int_0^{2\pi} d\theta \left(
   -\frac{\partial f}{\partial\epsilon} \right) \left( \cos\theta +
   \frac{v_2}{v_{\rm F}} \sin\theta \right)^2 ,
\end{equation}
where $\ell_0 = v_{\rm F} \tau(\bk_{\rm F} )$ is the quasiparticle
   mean free path at the nodes.
The final result crucially depends on the anisotropy ratio $v_2
   /v_{\rm F}$, and would be different in the two cases given by
   Eqs.~(\ref{eq:EkILT}) and (\ref{eq:Ekext}), due to their different
   $\theta$ dependence.
This has to be contrasted with the result obtained in the simple
   $d$-wave case, where~\cite{Krishana:97}:
\begin{equation}
\kappa_e = \eta k_{\rm B}^3 T^2 \frac{\ell_0}{v_2} \left( 1 +
   \frac{v_2^2}{v_{\rm F}^2} \right),
\end{equation}
with $\eta = (8\pi)^{-1} \int_0^\infty d x\, x^3 (-\partial
   f/\partial x)$.

\section{A limiting case}

In the absence of in-plane coupling, a spurious solution of the
   mean-field gap equation at $T=0$ can be
   implicitly expressed via~\cite{Angilella:99}:
\begin{equation}
E_\bk = \frac{1}{2} T_J (\bk) = \frac{1}{2} T_J g^4 (\bk).
\end{equation}
In such a limiting case, the superconducting energy spectrum would have
   purely kinetic origin, and would be identified with the interlayer
   pair-tunneling amplitude, divided by two.
A closed expression can then be obtained for ${\cal F}_1 (\beta)$, by
   utilizing the useful result:
\begin{equation}
\frac{1}{(2\pi)^2} \int d^2 \bk {\cal G} [\eta(\bk)] = \frac{2}{\pi^2} 
   \int_{-1}^1 dx {\cal G} (x) K(\sqrt{1-x^2}),
\end{equation}
where ${\cal G}[\eta(x)]$ is any continuous functional of $\eta(\bk) = 
   h(\bk)$ or $g(\bk)$ alone, and $K(x^\prime )$ ($x^\prime =
   \sqrt{1-x^2}$) is the complete elliptic integral of first
   kind~\cite{Gradshteyn:94}. 
From Eq.~(\ref{eq:integral}), expanding $K(x^\prime )$ around $x=0$,
   one eventually arrives at the closed
   expression:
\begin{equation}
{\cal F}_1 (\beta) \sim \frac{1}{\pi^2} \Gamma \left( \frac{1}{4} \right) 
   \frac{1}{\zeta^{1/4}} \left( 2\log 2 -\frac{1}{4} \psi \left(
   \frac{1}{4} \right) + \log \zeta^{1/4} \right),
\end{equation}
where $\psi(x)$ is the digamma function~\cite{Gradshteyn:94}, and the
   ILT amplitude $T_J$ itself here fixes the appropriate energy scale, 
   through $\zeta = \frac{1}{2} \beta T_J$.


\bibliography{r}
\bibliographystyle{prsty}

\end{document}